\documentclass[twocolumn,a4paper, showpacs, longbibliography, aps, pra, 10pt, amsmath, amssymb, superscriptaddress, preprintnumbers]{revtex4-1}
\usepackage{amsmath}  % needed for \tfrac, \bmatrix, etc.
\usepackage{amsfonts} % needed for bold Greek, Fraktur, and blackboard bold
\usepackage{amssymb}  % needed for symbols like \boxdot
\usepackage{graphicx} % needed for figures
\usepackage{upgreek}  % needed for non-italic greek letters
\usepackage[colorlinks,linkcolor=blue,citecolor=blue,urlcolor=blue]{hyperref}

\begin{document}

\title{\Large{Signatures of a dissipative phase transition in photon correlation measurements}}

\author{Thomas Fink} \email{finkt@phys.ethz.ch}
\affiliation{Institute of Quantum Electronics, ETH Z\"urich, CH-8093 Z\"urich, Switzerland}
\author{Anne Schade}
\affiliation{Technische Physik, Universit\"at W\"urzbug, D-97074 W\"urzburg, Germany}
\author{Sven H\"ofling}
\affiliation{Technische Physik, Universit\"at W\"urzbug, D-97074 W\"urzburg, Germany}
\affiliation{SUPA, School of Physics and Astronomy, University of St Andrews, St Andrews, KY16 9SS, United Kingdom}
\author{Christian Schneider}
\affiliation{Technische Physik, Universit\"at W\"urzbug, D-97074 W\"urzburg, Germany}
\author{Ata\c{c} \.Imamo\u{g}lu}
\affiliation{Institute of Quantum Electronics, ETH Z\"urich, CH-8093 Z\"urich, Switzerland}

% 03.67.Lx: Quantum computation
% 03.67.Hk: Quantum communication
% 73.21.La: Electron states and collective excitations in Quantum Dots
% 42.50.-p: Quantum optics
% 78.67.Hc:
% 42.50.Ex:
% 78.30.Fs:

\date{\today}

\maketitle

\textbf{Understanding and characterizing phase transitions in driven-dissipative systems constitutes a new frontier for many-body physics \cite{Hartmann06,Diehl08,Verstraete09,Diehl10,Tomadin11,Houck12,Boite13,Carmichael15,Fink17,Wilson16,Biondi16}. A generic feature of dissipative phase transitions is a vanishing gap in the Liouvillian spectrum \cite{Kessler12}, which leads to long-lived deviations from the steady-state as the system is driven towards the transition. Here, we show that photon correlation measurements can be used to characterize the corresponding critical slowing down of nonequilibrium dynamics. We focus on the extensively studied phenomenon of optical bistability in GaAs cavity-polaritons \cite{Baas04,Boulier14}, which can be described as a first-order dissipative phase transition \cite{Drummond80,Casteels16,Casteels17}. Increasing the excitation strength towards the bistable range results in an increasing photon-bunching signal along with a decay time that is prolonged by more than nine orders of magnitude as compared to that of low density polaritons. In the limit of strong polariton interactions leading to pronounced quantum fluctuations, the mean-field bistability threshold is washed out. Nevertheless, the scaling of the Liouvillian gap closing as thermodynamic limit is approached provides a signature of the emerging dissipative phase transition. Our results establish photon correlation measurements as an invaluable tool for studying dynamical properties of dissipative phase transitions without requiring phase-sensitive interferometric measurements.}

Bistability, the coexistence of two stable states with different photon number under the same driving conditions, is a general signature of driven-dissipative nonlinear systems, and is commonly observed in a broad range of experimental realizations \cite{Gibbs76,Dorsel83,Rempe91,Almeida04,Baas04,Notomi05,Wurtz06,Boulier14}. This phenomenon can be well understood by treating the system in a mean-field description, i.e. replacing the system's creation and annihilation operators by their corresponding expectation values. Neglecting quantum fluctuations in this way leads to diverging lifetimes of the two (otherwise metastable)  states and results in hysteretic behavior when the excitation power is ramped. A full quantum treatment, on the other hand, predicts a unique steady-state with no indication of the underlying bistability \cite{Drummond80,Kheruntsyan99}. These two pictures can be reconciled by investigating quantum nonequilibrium dynamics, described by a Liouvillian superoperator. In analogy with the spectrum of a Hamiltonian, the real part of the Liouvillian eigenvalues corresponds to the decay rate of the respective eigenmodes. The steady-state of the driven-dissipative evolution is the eigenmode of the Liouvillian with zero eigenvalue. In the bistable regime, the Liouvillian excitation gap, or the asymptotic decay rate of its first excited state, vanishes in the thermodynamic limit corresponding to a diverging lifetime of a second eigenmode. The latter can exceed all intrinsic time scales by many orders of magnitude and is the reason for the successful description of bistability experiments by mean-field theory. In general, a vanishing Liouvillian excitation gap signals a dissipative phase transition (DPT) \cite{Hioe81,Lett81,Kessler12,Casteels16,Casteels17,Letscher17}.

Here, we demonstrate that the asymptotic decay rate of the Liouvillian can be determined by measuring photon correlations in a Hanbury Brown and Twiss (HBT) set-up. Due to the coexistence of a high- and low-intensity state, detection of a first photon corresponds to a weak measurement of a high-photon number state, which in turn results in a higher probability to simultaneously detect a second photon compared to the average photon detection rate. This leads to a photon bunching signal, where the magnitude of bunching depends on the system nonlinearity and laser detuning. More importantly, the time scale over which the bunching persists depends on the decay rate towards the steady-state, and therefore reveals information about the Liouvillian gap (see Supplementary Information). We emphasize that the corresponding photon bunching may well exceed 2, indicating that photon correlations can be much stronger than those in thermal states of light.

\begin{figure}
\centerline{{\includegraphics[width=\linewidth]{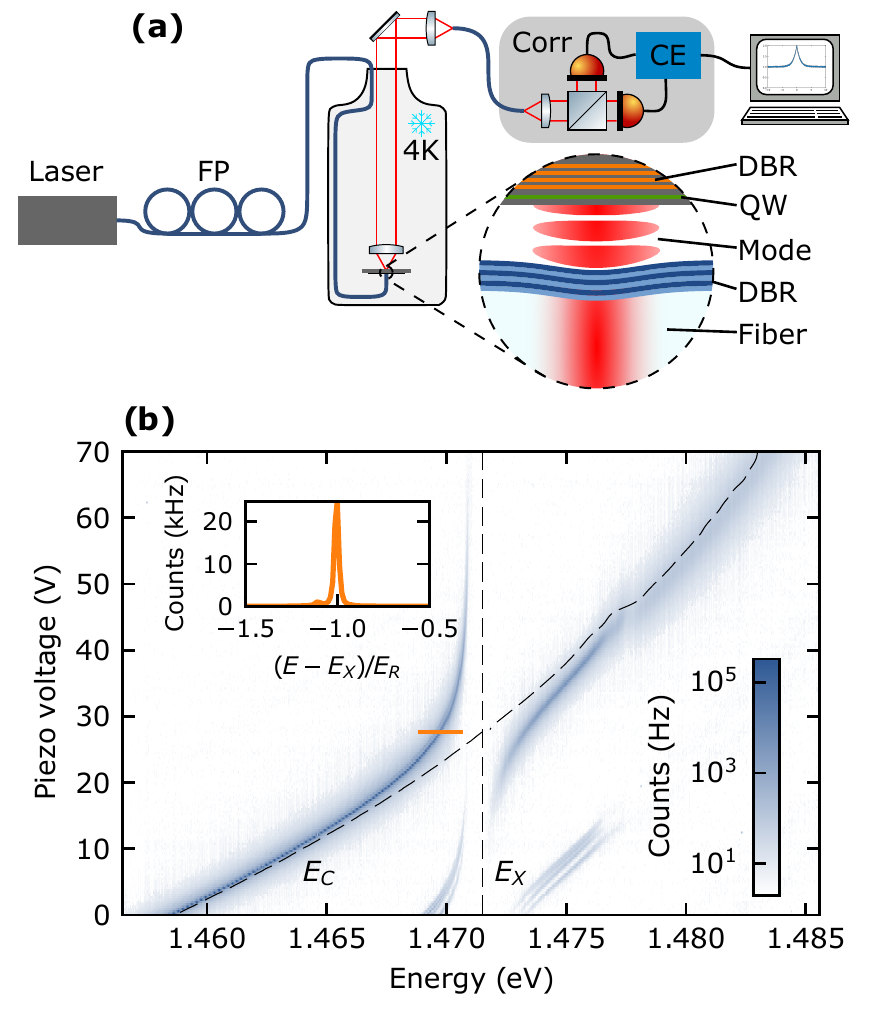}}}
\caption{\textbf{Experimental setup and cavity transmission spectrum.} (a) Intensity-stabilized cw laser light is coupled into the cavity via optical fibers (blue lines). Fiber paddles (FP) are used to control the polarization. The transmitted signal is directed onto the correlation setup (Corr). Two APDs in a HBT configuration record photon coincidences which are analyzed by correlation electronics (CE). The zoom-in displays the cavity structure. (b) Transmission spectrum under weak broadband illumination while sweeping the piezo voltage / cavity length. A clear anticrossing between the fundamental Gaussian cavity mode and exciton resonance (grey dashed lines denoted $E_{C,X}$) indicates the appearance of upper and lower polariton modes with a normal mode splitting of 3.50\,meV. The multiple resonances at low piezo voltages correspond to the TEM$_{01,10}$ transverse mode manifold. At 1.478\,eV another anticrossing with smaller splitting appears which may stem from strong coupling of the cavity mode to an impurity resonance in the semiconductor. Inset: linecut along the orange line showing the lower polariton resonance. A small linear polarization splitting of 180\,$\upmu$eV is visible.}
\label{fig1}
\end{figure}

Figure~\ref{fig1}~(a) shows the schematic of the experiment where a fiber cavity incorporating an InGaAs quantum well (QW) is driven by a single-mode laser field (see Methods). Application of a dc voltage to a nanopositioner that controls the distance between fiber and sample allows for in-situ tuning of the cavity energy with respect to the QW exciton resonance. Sweeping this voltage while recording the transmission spectrum under broadband illumination in Fig.~\ref{fig1}~(b) results in an anticrossing demonstrating the strong coupling between cavity photons and QW excitons. The new eigenstates of the coupled system are the lower and upper polaritons. The large splitting between the fundamental longitudinal TEM$_{00}$ and next transverse TEM$_{10,01}$ modes of 11.8\,meV illustrates the strong transverse confinement generated by the curvature of the fiber DBR. Driving the TEM$_{00}$ mode of the lower polariton branch near-resonantly constitutes a realization of a driven-dissipative system with Kerr-type nonlinear interactions. The corresponding Hamiltonian in a polariton basis rotating with the excitation laser frequency is
\begin{equation} \label{eq:ham}
\hat{H} = \Delta \hat{a}^\dagger \hat{a} + \frac{U}{2} \hat{a}^\dagger \hat{a}^\dagger \hat{a} \hat{a} + F^* \hat{a}^\dagger + F \hat{a},
\end{equation}
where $\hat{a}$ is a polariton annihilation operator, $\Delta = E_L - E_P$ is the detuning between excitation laser and polariton mode energy, $U$ is the single-particle polariton-polariton interaction, and $F$ is the laser drive strength. The nonlinearity originates from exciton-exciton interactions weighted by the exciton content of the polariton and enhanced by the strong transverse confinement. By changing the cavity-exciton detuning, we can change the lower polariton mode from a purely photonic linear cavity mode to an excitonic one with sizable nonlinear interactions.
Since polaritons decay at a rate $\gamma$, determined primarily by cavity mirror losses, the polariton dynamics obey the master equation
\begin{equation} \label{eq:master}
\frac{\partial \hat{\rho} }{\partial t} = \hat{\mathcal{L}} \hat{\rho},
\end{equation}
where
\begin{equation} \label{eq:liou}
\hat{\mathcal{L}} \hat{\rho} = \frac{1}{i \hbar} [\hat{H},\hat{\rho}] + \frac{\gamma}{2} \left( 2 \hat{a} \rho \hat{a}^\dagger - \hat{a}^\dagger \hat{a} \rho - \rho \hat{a}^\dagger \hat{a} \right)
\end{equation}
is the Liouvillian superoperator with $\gamma$ denoting the polariton decay rate. Nonlinear driven-dissipative polariton systems described by Eq.~\ref{eq:master} not only constitute a platform to test fundamental quantum optics phenomena \cite{Kasprzak06,Carusotto13}, but are also potential candidates for applications in quantum simulations of nonequilibrium dynamics \cite{Jacqmin06,Liew08,Amo10,Gao12,Ballarini13,Schneider17}.

\begin{figure}
\centerline{{\includegraphics[width=\linewidth]{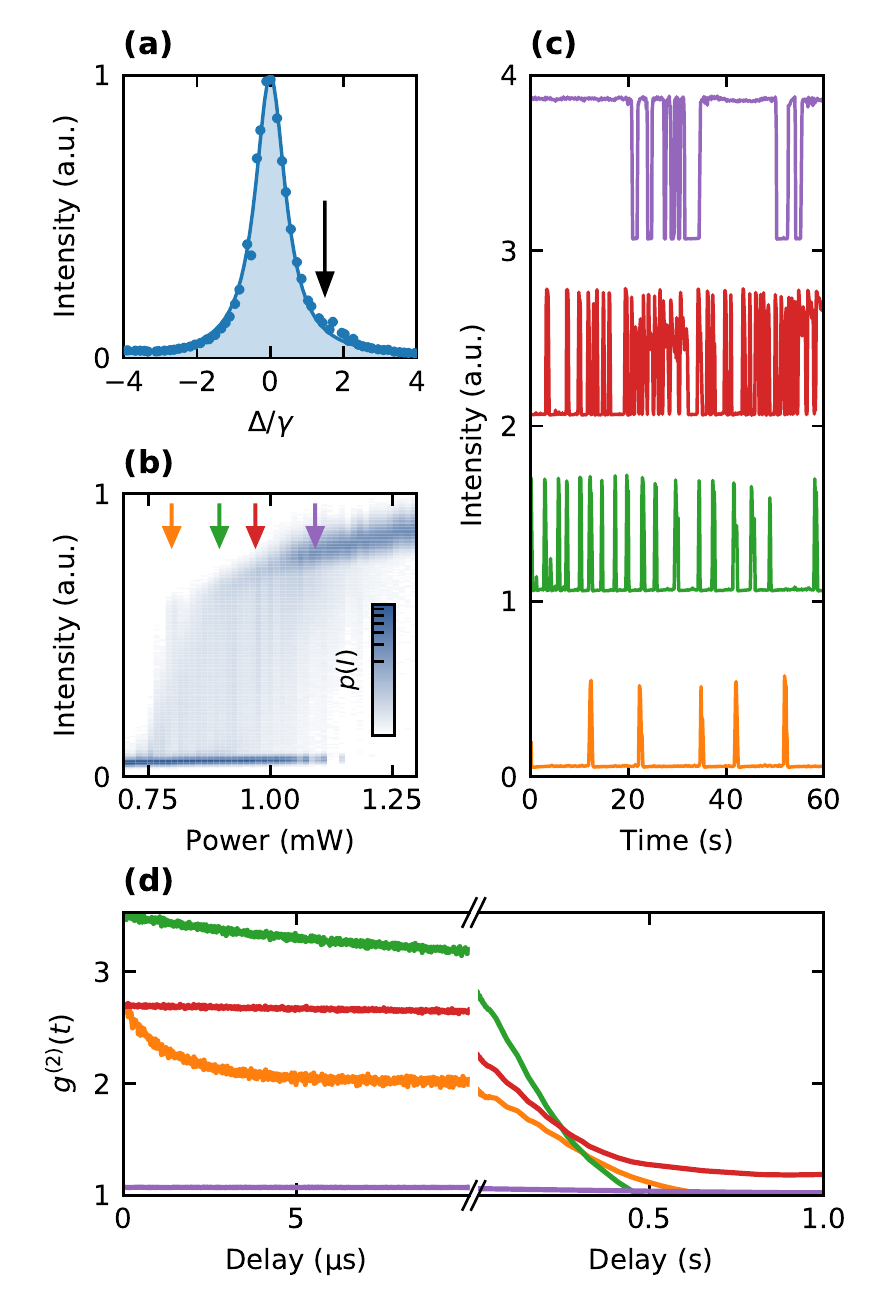}}}
\caption{\textbf{Dynamical bistability in cavity transmission.} (a) Normalized polariton transmission spectrum under weak excitation (100\,nW) as a function of the laser detuning. The blue solid line is a Lorentzian fit to the data (blue circles) with a mode linewidth (FWHM) of $\gamma=37\,\upmu$eV. The black arrow denotes the detuning of the excitation laser with respect to the polariton mode $\Delta / \gamma = 1.5$ used in (b)-(d). (b) Color-coded photon number distribution $p(I)$ obtained from 150\,s-long photon traces with 1\,ms time binning. Between 750 and 1,200\,$\upmu$W, the distribution is double-peaked indicating the bistable regime. Note the nonlinear color scaling used to highlight the small intensity noise in between that occurs due to switching events faster than the binning time. The arrows denote the powers where the data in (c) and (d) has been taken. (c) Time-binned (100\,ms) single photon count events normalized to the maximum intensity of all traces showing fluctuation-induced switching between the high- and low-intensity state. (d) $g^{(2)}$ function obtained from the data in (c). Clear photon bunching persisting up to the second scale is observed. For delays longer than 1\,ms, classical intensity autocorrelations (see Methods) are shown. Note the discontinuous time axis showing both short (0-10\,$\upmu$s) and long dynamics up to 1\,s.}
\label{fig2}
\end{figure}

To investigate the dynamical signatures of dispersive bistability, we excite the polariton mode with a continuous-wave laser at a detuning of $\Delta / \gamma=1.5$ (see Fig.~\ref{fig2}~(a)). In this regime (i.e. for $\Delta / \gamma \geq \sqrt{3} /2$), a classical description of the system has two solutions and mean-field theory predicts bistability \cite{Drummond80,Baas04}. Indeed, when varying the power in Fig.~\ref{fig2}~(b) we observe a region where the intensity probability distribution, obtained by recording photon traces for 150\,s and calculating the statistical distribution of photon counts in 1\,ms wide time bins, is peaked around two values with small and large intensity, respectively. Recording individual photon detections then reveals the system dynamics illustrated in Fig.~\ref{fig2}~(c): We observe random switching events between a low- and high-intensity state triggered by fluctuations in the system. The time scale of these events on the order of hundreds of ms corresponds to a critical slowing down by more than nine orders of magnitude compared to the bare cavity lifetime, which we measured to be 35\,ps in a pulsed excitation experiment independently. This dramatic slowing down is a striking and direct signature of a DPT.

The second-order correlation function
\begin{equation} \label{eq:g2}
g^{(2)}(t) = \frac{\left\langle \hat{a}^\dagger (t') \hat{a}^\dagger (t' + t) \hat{a} (t' + t) \hat{a} (t') \right\rangle}{\left\langle \hat{a}^\dagger (t') \hat{a} (t') \right\rangle^2},
\end{equation}
allows for determining the dynamical properties of the polariton mode. In a nut shell, $g^{(2)}(t)$ yields the  dynamics of the polariton system conditioned upon detection of a photon at time $t'$. Since the collapse operator $\hat{a}$ does not commute with $\hat{H}$, detection of a photon projects the nonlinear system out of its steady-state into a mixture of several Liouvillian eigenstates. The conditional dynamics following the initial photon detection are interrupted by the second photon detection event at time $t'+t$, which effectively determines polariton relaxation dynamics towards the steady-state. In this sense, $g^{(2)}(t)$ measures the dissipative counterpart of Hamiltonian quench dynamics.

Figure~\ref{fig2}~(d) shows photon correlation measurements obtained for the pump intensities indicated in Fig.~\ref{fig2}~(b), where we observe a biexponential decay with two distinct time scales. The decay on $\upmu$s timescale strongly depends on the pump strength and we tentatively attribute this decay rate to the real part of the second excited Liouvillian eigenstate, which can also undergo a reduction as a function of drive strength (see Supplementary Information). In stark contrast, the decay of correlations in the ms timescale is largely independent of the pump power. This observation suggests that the physics at these timescales is no longer determined by the Liouvillian of Eq.~\ref{eq:liou} and that the influence of external classical noise sources, such as acoustic noise affecting the fiber cavity, needs to be taken into account.

\begin{figure*}
\centerline{{\includegraphics[width=\linewidth]{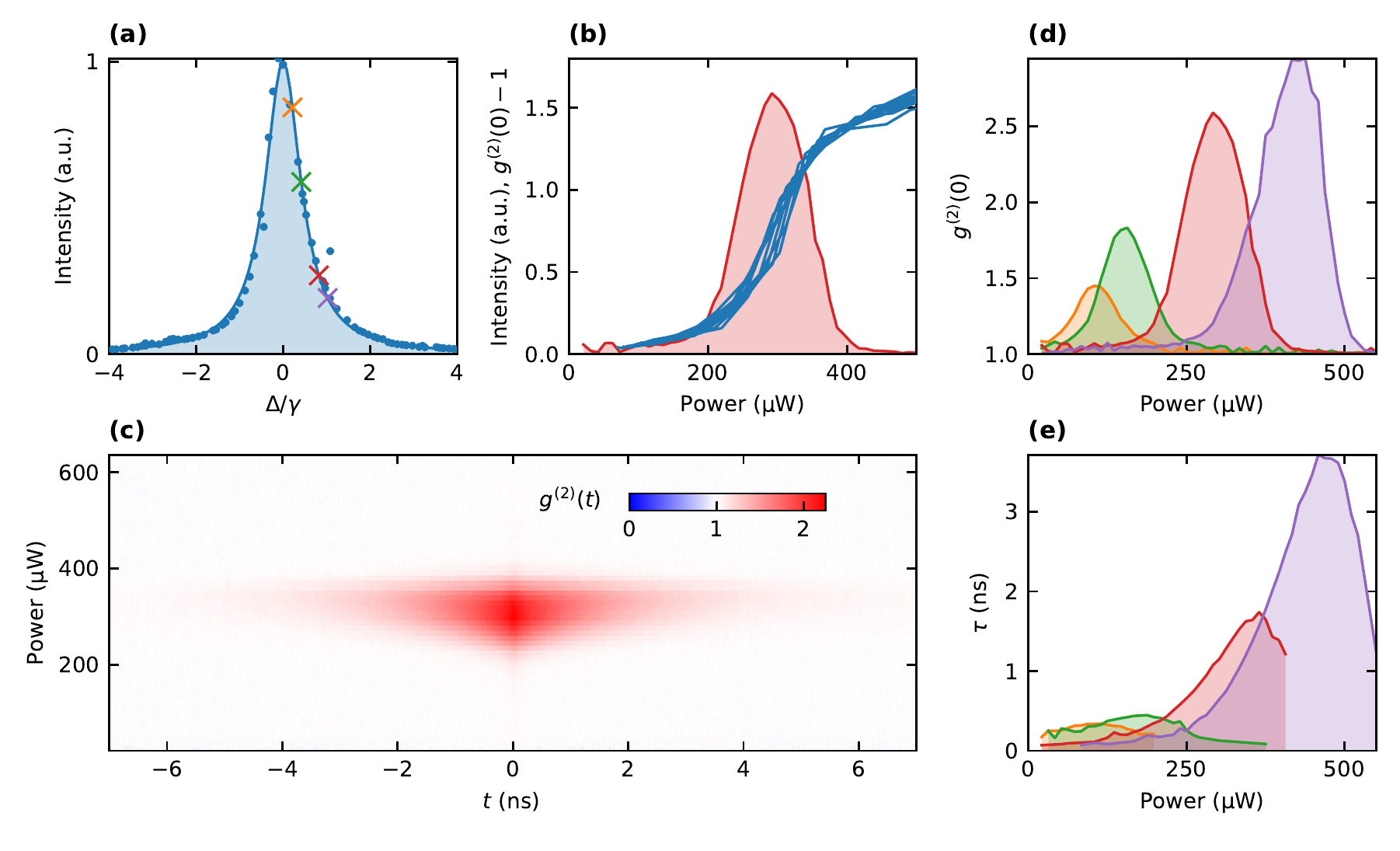}}}
\caption{\textbf{Nonequilibrium dynamics in the strong fluctuation regime.} (a) Normalized polariton transmission spectrum under weak laser drive power of 100\,nW as a function of laser detuning. The blue solid line is a Lorentzian fit to the data (blue dots). The crosses indicate the laser detunings $\Delta / \gamma = 0.2,\,0.4,\,0.8,$ and 1.0, where the data in (b)-(e) has been taken. (b) Blue line: transmitted intensity while ramping the excitation power at $\Delta / \gamma = 0.8$ (red cross in (a)) up and down repeatedly. No bistability but only a single trace with superlinear behavior is visible. Red line: bunching amplitude $g^{(2)}(0)-1$ for the same detuning (same data as red line in (d)). Strongest bunching coincides with the onset of superlinear behavior in a power ramp. (c) Power-dependent $g^{(2)}(t)$ function for $\Delta / \gamma =0.8$. The slowing down of dynamics is observable through the prolongation of photon bunching at the critical drive strength. (d) and (e) Bunching amplitude and time scale extracted from power-dependent $g^{(2)}$ measurements for the detunings indicated in (a).}
\label{fig3}
\end{figure*}

Whereas for detunings exceeding $\sqrt{3}/2\,\gamma$ mean-field theory successfully predicts the bimodal intensity distribution observed in Fig.~\ref{fig2}~(b), only a single solution is obtained for smaller values and one thus expects simple dynamics. To explore if nonequilibrium dynamics exhibit a sharp detuning dependence in the presence of quantum fluctuations, we increase the polariton nonlinearity by increasing the exciton content of the lower polariton branch and change the laser detuning at which we excite the polariton mode (Fig.~\ref{fig3}~(a)). In Fig.~\ref{fig3}~(b) we show that bunching is strongest for excitation powers at the nonlinear threshold where a power ramp exhibits a superlinear behavior and strong photon number fluctuations occur. Above and below this region, photon statistics converge to the coherent state value $g^{(2)}(0)=1$ (Fig.~\ref{fig3}~(c)).

Remarkably, we observe that changing the laser detuning from the stable ($\Delta / \gamma < \sqrt{3}/2$) to the bistable ($\Delta / \gamma > \sqrt{3}/2$) regime does not result in a qualitative change in either the bunching amplitude or the decay time (Fig.~\ref{fig3} (d) and (e), respectively). The closing of the Liouvillian gap is more pronounced for larger laser detunings where the mean polariton number required to reach the onset of nonlinear response is larger and hence longer-lived bunching occurs (see Supplementary Information). However, the sharp distinction of single and multiple solutions in mean-field theory as a function of laser detuning is washed away by quantum fluctuations that are enhanced by the strong polariton interactions which we estimate to be $U / \gamma \sim 0.03$.

The decay times of the $g^{(2)}(t)$ bunching peak depicted in Fig.~\ref{fig3} are orders of magnitude shorter than those in Fig.~\ref{fig2} due to the enhanced nonlinearity and reduced mean polariton number. Naturally, with a slow laser power scan we do not observe bistability even for  $\Delta / \gamma = 1.0 > \sqrt{3}/2$. On the other hand, a lack of a qualitative differences in bunching and Liouvillian gap closing across the "mean-field bistability threshold" implies the possibility to observe bistability even for $\Delta / \gamma < \sqrt{3}/2$ if the laser power is ramped up/down faster than the inverse Liouvillian gap.

\begin{figure*}
\centerline{{\includegraphics[width=\linewidth]{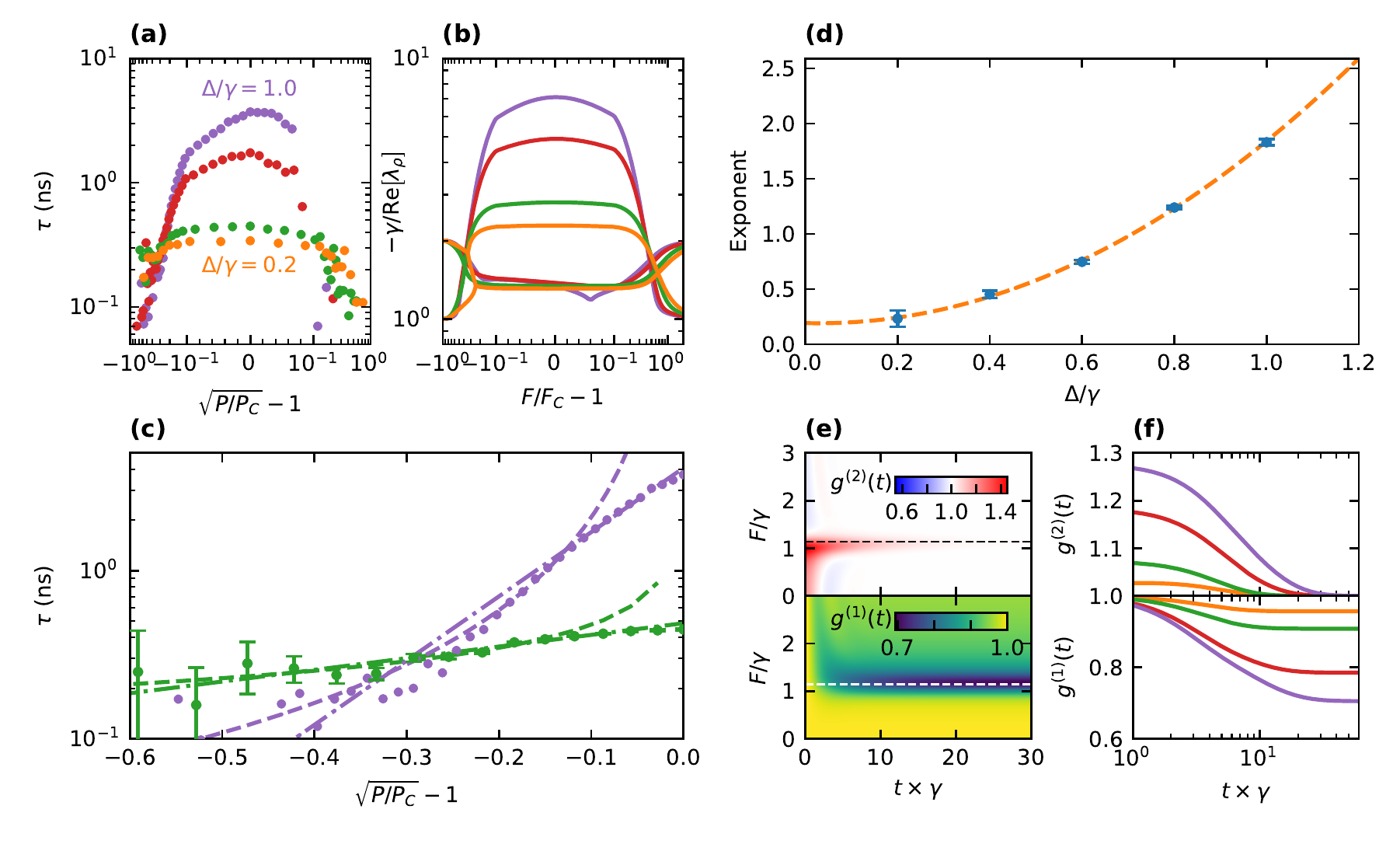}}}
\caption{\textbf{Liouvillian gap scaling with drive strength.} (a) Bunching timescale in the vicinity of the critical drive strength corresponding to the longest observed bunching decay time for each detuning. Different line colors denote detunings $\Delta / \gamma = 0.2,\,0.4,\,0.8,$ and 1.0 (color code same as in Fig.~\ref{fig3}). (b) Simulation of the Liouvillian excited state lifetime for the same detunings and $U / \gamma = 0.2$ showing good qualitative agreement with the data in (a). (c) Zoom-in on the region where $\sqrt{P/P_C} - 1 < 0$ in (a). Dashed lines and dash-dotted lines denote weighted power-law and exponential fits, respectively. For clarity, only detunings 0.4 (green) and 1.0 (purple) are shown, and error bars (one standard deviation) are displayed only for the green curve. (d) Exponent of the power-law fits in (c) as a function of laser detuning. Error bars are one standard deviation. The orange dashed line is a weighted quadratic fit. (e) Simulations of power- and time-dependent $g^{(2)}$ and $g^{(1)}$ functions (top and bottom panel, respectively). Both display qualitative changes when the critical drive strength is reached. The detuning is $\Delta / \gamma = 1.0$ and other parameters are the same as in (b). (f) Time-dependent $g^{(2)}$ and $g^{(1)}$ functions (top and bottom panel, respectively) for excitation powers where the longest-lived bunching occurs (linecuts along dashed lines in (e)) for the same detunings as in (a)-(c).}
\label{fig4}
\end{figure*}

Even though the bunching amplitude and the Liouvillian gap in Fig.~\ref{fig3} (d)-(e) show a monotonous quantitative change across the bistability threshold for increasing laser detuning, their scaling behaviour as a function of drive strength reveals the signatures of the underlying DPT. Despite the fact that we are dealing with a first-order transition in a zero-dimensional system that is far from the thermodynamic limit, we follow Ref.~\cite{Casteels17} and refer to the drive strength for which the longest bunching is observed as the critical drive $F_C=\sqrt{P_C}$. In Fig.~\ref{fig4}~(a) we show the bunching decay timescale as a function of normalized distance to this point. Whereas for second-order phase transitions a dependence $\tau \sim \left| F/F_C -1 \right|^\alpha$ with a universal exponent $\alpha$ and divergence at the critical point would be expected \cite{Sachdev11}, no such behavior is known for first-order transitions. Indeed, we find a laser detuning-dependent finite Liouvillian excited state lifetime $-\gamma / \mathrm{Re} \left[ \lambda_\rho \right]$ around the critical drive, in good qualitative agreement with numerical simulations (Fig.~\ref{fig4}~(b)). In Ref.~\cite{Casteels17}, a power-law behavior was suggested to describe the lifetime scaling when approaching the critical drive. In Fig.~\ref{fig4}~(c), we indeed find good agreement between a power-law fit (dashed lines) and the data in a limited region away from the critical power. We emphasize, however, that using an exponential fit also provides a reasonable description of the whole curve until the critical point (dash-dotted lines).

For a given polariton-polariton interaction strength, the mean polariton number required to reach the bistable regime increases linearly with $\Delta/\gamma$, indicating that increasing the laser detuning suppresses the effect of quantum fluctuations. The increase of the power-law exponent with increasing $\Delta/\gamma$ in Fig.~\ref{fig4}~(d) shows that as we in this fashion push the system towards the thermodynamic limit, the Liouvillian gap will close abruptly as the classically bistable region is reached.

The information about the Liouvillian spectrum of driven-dissipative systems could also be obtained using first-order coherence ($g^{(1)}$) measurements. In Fig.~\ref{fig4}~(e) we plot calculated $g^{(2)}$ and $g^{(1)}$ as a function of time and drive strength. Indeed, both exhibit long-lived deviations from an all-orders coherent response at the critical drive strength and the bunching observed in $g^{(2)}$ corresponds to a decrease of coherence in $g^{(1)}$. Line-cuts along the dashed lines, where the longest-lived bunching is observed, are displayed in Fig.~\ref{fig4}~(f) and demonstrate that the decay of bunching and coherence are dictated by the same time scale. For the long dynamics observed here, however, a $g^{(1)}$ measurement is experimentally much harder to realize, and may even be infeasible: Since interference experiments are sensitive to phase noise, it is highly challenging to introduce sufficiently large delays to probe second-scale dynamics and to ensure phase coherence of the excitation light over such long times. $g^{(2)}$ measurements, on the other hand, are insensitive to phase coherence and are therefore superior from an experimental point of view for problems involving such large dynamic ranges as reported here.

We conclude that measuring $g^{(2)}$ of driven-dissipative systems under near-resonant excitation reveals crucial insights into the underlying nonequilibrium dynamics described by the Liouvillian eigenspectrum. Moreover, this technique is applicable to a whole range of systems from close to thermodynamic limit down to the highly interesting regime where quantum fluctuations are prominent. Whereas dynamical hysteresis also allows to extract the Liouvillian gap \cite{Rodriguez17}, the requirement to sweep the excitation power nonadiabatically restricts these measurements to regimes close to the thermodynamic limit with vanishing gap and diverging timescales. Applying our $g^{(2)}$-based technique to lattices of interacting photonic systems could pave the way to study the rich field of nonequilibrium quantum many-body dynamics, which recently garnered a lot of interest due to their intriguing steady-state phase diagrams \cite{Tomadin10,Boite13,Jin14,Wilson16,Mendoza16,Moss17}.

\newpage
\noindent {\fontfamily{phv}\selectfont
\textbf{Methods} }

\noindent\textbf{Setup.}
The experimental setup is illustrated in Fig.~\ref{fig1}~(a). The sample consists of an epitaxially grown GaAs $\lambda$ spacer layer containing a single 17\,nm-wide In$_{0.04}$Ga$_{0.96}$As quantum well on top of 34 pairs of AlAs / GaAs distributed Bragg reflectors. It faces a dielectric DBR with 17.5 pairs deposited on the facet of an optical fiber. A Gaussian dimple-shaped surface geometry has been fabricated onto the facet by CO$_2$ laser ablation prior to coating, such that in addition to the longitudinal confinement of the cavity mode due to the DBRs also strong transverse confinement is induced resulting in a zero-dimensional polariton box with discrete eigenstates \cite{Besga15}. Due to a finite ellipticity of the cavity the photonic mode is split by 180\,$\upmu$eV into two linearly polarized modes out of which only one is addressed by the excitation through careful alignment of the polarization of the incoming light. The transmitted signal is sent to the top of a dewar in a free-space configuration. The sample is mounted in a bath cryostat at 4\,K temperature.\\

\noindent\textbf{Measurements.}
For excitation, a tunable single-mode diode laser is used. Before being coupled into the setup, the light passes a noise eater to reduce laser intensity noise. The power values mentioned in the main text correspond to the power sent into the setup, whereas due to mode mismatch between the plane waves in the fiber core and the intracavity mode, as well as an asymmetric DBR configuration, the incoupling efficiency into the cavity is on the order of $10^{-4}-10^{-3}$. The transmitted light is then coupled into an optical fiber and sent to two APDs in a HBT configuration to measure both transmission spectra and photon statistics. The time resolution (instrument response function full width at half maximum) is 64\,ps.\\

\noindent\textbf{Photon correlations.}
In Fig.~\ref{fig2}, single photon coincidences have been recorded in the so-called time-tagged time-resolved (TTTR) mode of the correlation electronics. This allows us to obtain both time traces and $g^{(2)}$ from the data. To analyze them, we employ different techniques to calculate $g^{(2)}(t)$. For small delays up to 10\,ms, we calculate the true single-photon correlation function as defined in Eq.~\ref{eq:g2} with a delay bin width of 400\,ps. Then, we downsample this data to 10\,ns to obtain the curves displayed in Fig.~\ref{fig2}~(d). To restrict the computation time required to treat longer delays, we then bin all photon arrival times into 10\,$\upmu$s windows and calculate the classical intensity autocorrelation function $g^{(2)}_{class}(t)=\left\langle I(t') I(t'+t) \right\rangle / \left\langle I(t') \right\rangle^2$ for longer delays. For the data in Fig.~\ref{fig3} and \ref{fig4} we directly process and save photon correlation data without recording individual arrival times. To extract the bunching amplitude and time scale we fit an exponentially decaying bunching signature $g^{(2)}(t)=1+A \exp\left( - \left| t \right | / \tau \right)$ convolved with the independently measured instrument response function to the data.

%\bibliography{dpt}


\vspace{5mm}
\noindent {\large \textbf{Acknowledgments}}

\noindent We would like to thank A. Reinhard, T. Volz, and J. Reichel for early work that lead to the development of the semiconductor fiber cavity structure used in this work. We also acknowledge fruitful discussions with C. Ciuti and S. Zeytino\u{g}lu. This work was supported by the Swiss National Science Foundation (SNSF) through the National Centre of Competence in Research - Quantum Science and Technology (NCCR QSIT). A.S., C.S., and S.H. acknowledge support by the State of Bavaria and the DFG within the Project Schn1376/3-1. \\

\noindent {\large  \textbf{Contributions}}

\noindent T.F. and A.\.I. designed and supervised the experiment. T.F. carried out the measurements. A.S., S.H., and C.S. grew the sample. T.F. and A.\.I. wrote the manuscript.\\

\noindent {\large \textbf{Competing financial interests}}

\noindent The authors declare no competing financial interests.\\

\noindent {\large  \textbf{Corresponding authors}}

\noindent Correspondence to: Thomas Fink or Ata\c{c} \.Imamo\u{g}lu.\\

\clearpage
\widetext

\renewcommand{\thefigure}{S\arabic{figure}}
\setcounter{figure}{0}

\begin{center}
\large{\textbf{Supplementary Information}}
\end{center}

\begin{figure*}
\centerline{{\includegraphics[width=\linewidth]{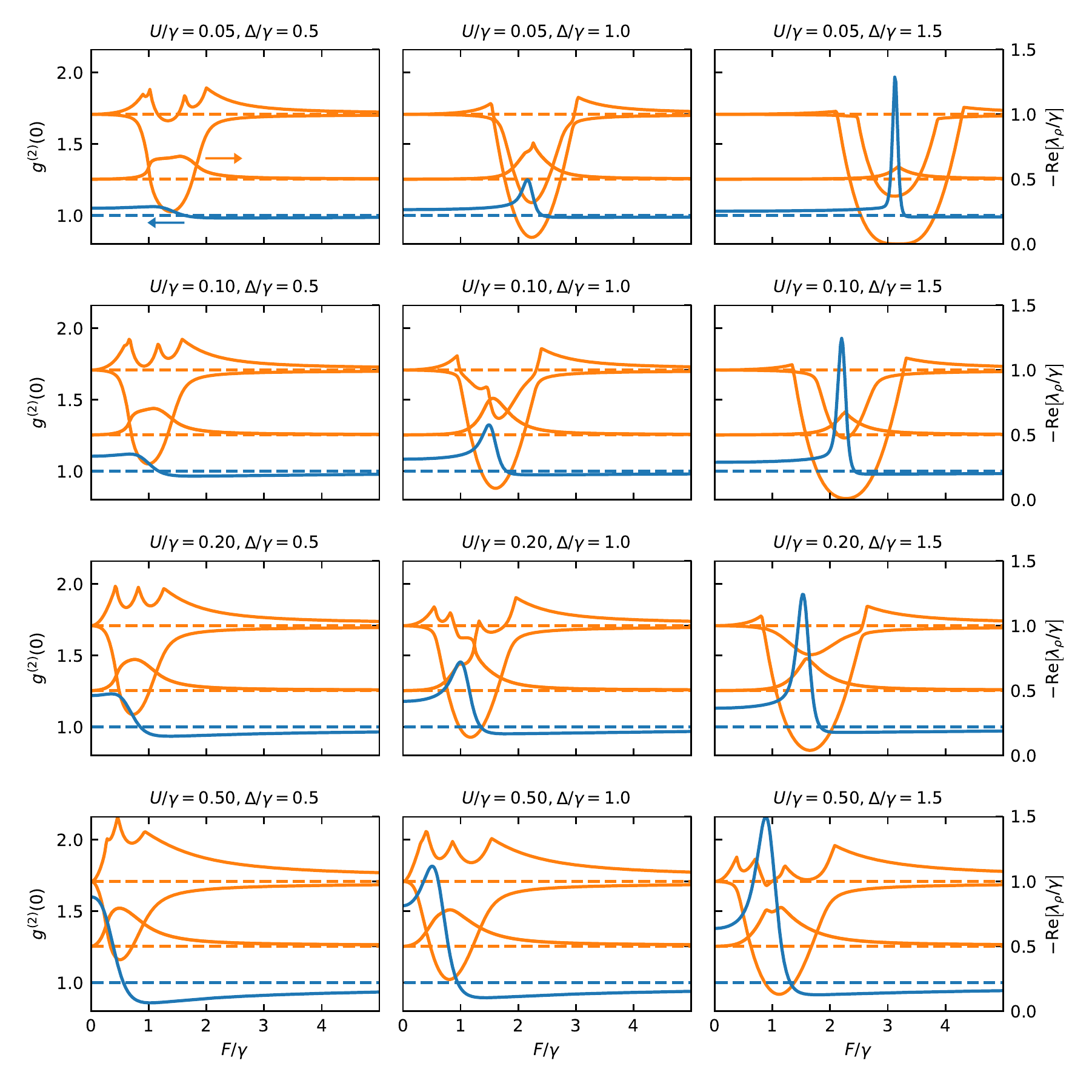}}}
\caption{\textbf{Bunching and asymptotic decay rate.} Steady-state bunching amplitude (solid blue line, left axis) and lowest Liouvillian excited states decay rates (solid orange lines, right axis) as a function of normalized drive strength $F / \gamma$ for increasing nonlinearity (top to bottom) and laser detuning (left to right). The respective values for a linear system are indicated by dashed lines. For small nonlinearities and large detunings, the decay rate vanishes and hence the system time scale diverges. Increasing the nonlinearity while keeping the detuning constant increases the bunching amplitude but reduces the Liouvillian excited state lifetime, whereas increasing the detuning while keeping the nonlinearity constant increases both bunching amplitude and lifetime. The Hilbert space has been truncated at the $N=90$ manifold where convergence was ensured.}
\label{figadr}
\end{figure*}

The dynamics of the driven-dissipative nonlinear exciton-polariton system is described by Eqs.~\ref{eq:ham}-\ref{eq:liou} of the main text. To gain insights into the behavior of photon correlation signatures while changing excitation laser detuning and system nonlinearity, we calculate the steady-state bunching amplitude and the Liouvillian excitation spectrum with corresponding eigenvalues $\lambda_\rho$ (Fig.~\ref{figadr}).

\end{document}